\input form
\def\Oscr{{\cal O}}
\cita{Arnold-1963}{V.I. Arnold: {\it  Proof of a theorem of A. N.
Kolmogorov on the invariance of quasi-periodic motions under small
perturbations of the Hamiltonian}, Usp. Mat. Nauk, {\bf 18}, 13 (1963);
Russ. Math. Surv., {\bf 18}, 9 (1963).}
\cita{Arnold-1963.1}{V.I. Arnold: {\it  Small denominators and problems
of stability of motion in classical and celestial mechanics},
Usp. Math. Nauk {\bf 18} N.6, 91 (1963); Russ. Math. Surv., {\bf 18}
N.6, 85 (1963).}
\cita{Carpino-1987}{M. Carpino, A. Milani, A. Nobili: {\it  Long term
numerical integration and synthetic theories for the motion of outer
planets}, Astronomy and Astrophysics, {\bf 181}, 182--194 (1987).}
\cita{Celletti-1991}{A. Celletti and A. Giorgilli: {\it  On the stability
of the Lagrangian points in the spatial restricted problem of three
bodies}, Celest. Mech. Dyn. Astr., {\bf 50}, 31--58 (1991).}
\cita{Celletti-2000}{Celletti, A., Giorgilli, A. and Locatelli, U.:
{\it  Improved Estimates on the Existence of Invariant Tori for Hamiltonian
Systems}, Nonlinearity, {\bf 13}, 397--412 (2000).}
\cita{Celletti-2005}{A. Celletti, L. Chierchia: {\it  KAM stability and
celestial mechanics},  Mem. Amer. Math. Soc., {\bf 187}, 1--134 (2007).}
\cita{Contopolus-2003}{G. Contopoulos, C. Efthymiopoulos and A. Giorgilli:
{\it  Non-convergence of formal integrals of motion}, J. Phys. A:
Math. Gen., {\bf 36}, 8639--8660 (2003).}
\cita{Contopolus-2004}{C. Efthymiopoulos, G. Contopoulos and A. Giorgilli:
{\it  Non-convergence of formal integrals of motion II: Improved
Estimates for the Optimal Order of Truncation}, J. Phys. A:
Math. Gen., {\bf 37}, 10831--10858 (2004).}
\cita{Efthimiopoulos-2005}{C. Efthimiopoulos, Z. S\'andor: {\it 
Optimized Nekhoroshev stability estimates for the Trojan asteroids
with a symplectic mapping model of co-orbital motion},
Mon. Not. R. Astron. Soc., {\bf 364}, 253--271 (2005).}
\cita{Efthimiopoulos-2008}{C. Efthimiopoulos: {\it 
On the connection between the Nekhoroshev theorem and Arnold diffusion},
Celestial Mechanics and Dynamical Astronomy, {\bf 102}, 49--68 (2008).}
\cita{Fejoz-2005}{J. Fejoz: {\it  D\'emonstration du ``th\'eor\`eme
d'Arnold'' sur la stabilit\'e du syst\`eme plan\'etaire (d'apr\`es
Michael Herman)}, Ergodic Theory Dyn. Sys., {\bf 24} N.5, 1521--1582 (2005).}
\cita{Gabern-2005}{F. Gabern, A. Jorba and U. Locatelli: {\it  On
the construction of the Kolmogorov normal form for the Trojan
asteroids}, Nonlinearity, {\bf 18} N.4, 1705--1734 (2005).}
\cita{Giorgilli-1989}{A. Giorgilli, A. Delshams, E. Fontich, L. Galgani, and
C. Sim\'o: {\it  Effective stability for a Hamiltonian system near an
elliptic equilibrium point, with an application to the restricted three
body problem.} J. Diff. Eqs., {\bf 77}, 167--198 (1989).}
\cita{Giorgilli-1997}{A. Giorgilli, Ch. Skokos: {\it  On the stability of
the Trojan asteroids}, Astronomy and Astrophysics, {\bf 317}, 254--261 (1997).}
\cita{Giorgilli-1997.5}{A. Giorgilli and A. Morbidelli: {\it  Invariant
KAM tori and global stability for Hamiltonian systems}, ZAMP, {\bf 48},
102--134 (1997).}
\cita{Giorgilli-1997.1}{A. Giorgilli and U. Locatelli, {\it  Kolmogorov
theorem and classical perturbation theory}, J. of App. Math. and
Phys. (ZAMP), {\bf 48}, 220--261 (1997).}
\cita{Gio-Loc-1997.1}{A. Giorgilli and U. Locatelli: {\it  On
classical series expansion for quasi-periodic motions}, MPEJ, {\bf
3} N.5, 1--25 (1997).}
\cita{Giorgilli-03.1}{A. Giorgilli: {\it  Notes on exponential stability
of Hamiltonian systems}, in {\it Dynamical Systems, Part I:
Hamiltonian systems and Celestial Mechanics} Pubblicazioni del Centro
di Ricerca Matematica Ennio De Giorgi, Pisa, 87--198 (2003).}
\cita{Kolmogorov-1954}{A.N. Kolmogorov: {\it  Preservation of
conditionally periodic movements with small change in the Hamilton
function}, Dokl. Akad. Nauk SSSR, {\bf 98}, 527
(1954). Engl. transl. in: Los Alamos Scientific Laboratory translation
LA-TR-71-67; reprinted in: Lecture Notes in Physics {\bf 93}.}
\cita{Laskar-1989}{J. Laskar: {\it  A numerical experiment on the
chaotic behaviour of the solar system}, Nature, {\bf 338}, 237--238
(1989).}
\cita{Laskar-1994}{J. Laskar: {\it  Large scale chaos in the solar
system}, Astronomy and Astrophysics, {\bf 287}, L9--L12 (1994).}
\cita{Laskar-99}{J. Laskar: {\it  Introduction to frequency map analysis},
in C. Sim\`o (managing ed.), Proceedings of the NATO ASI school:
``Hamiltonian Systems with Three or More Degrees of Freedom'', S'Agaro
(Spain), June 19--30, 1995, Kluwer, 134--150 (1999).}
\cita{Lhotka-2008}{C. Lhotka, C. Efthimiopoulos, R. Dvorak: {\it 
Nekhoroshev stability at L4 or L5 in the elliptic-restricted
three-body problem --- application to the Trojan asteroids},
Mon. Not. R. Astron. Soc., {\bf 384}, 1165--1177 (2008).}
\cita{Littlewood-1959}{J.E. Littlewood: {\it  On the equilateral
configuration in the restricted problem of three bodies}, Proc. London
Math. Soc.(3), {\bf 9}, 343--372 (1959).}
\cita{Littlewood-1959.1}{J.E. Littlewood: {\it  The Lagrange
configuration in Celestial Mechanics}, Proc. London Math. Soc.(3),
{\bf 9}, 525--543 (1959).}
\cita{Loc-Gio-2000}{Locatelli, U. and Giorgilli, A., {\it  Invariant tori in
the secular motions of the three-body planetary systems},
Celestial Mechanics and Dynamical Astronomy, {\bf 78}, 47--74 (2000).}
\cita{Loc-Gio-2005}{U. Locatelli and A. Giorgilli: {\it  Construction
of the Kolmogorov's normal form for a planetary system}, Regular and
Chaotic Dynamics, {\bf 10} N.2, 153--171 (2005).}
\cita{Locatelli-2007}{U. Locatelli, A. Giorgilli: {\it  Invariant tori
in the Sun--Jupiter--Saturn system}, DCDS-B, {\bf 7}, 377--398 (2007).}
\cita{Morbidelli-1995}{A. Morbidelli and A. Giorgilli: {\it 
Superexponential stability of KAM tori}, J. Stat. Phys., {\bf 78},
1607--1617 (1995).}
\cita{Moser-1955}{J. Moser: {\it  Stabilit\"atsverhalten kanonisher
differentialgleichungssysteme}, Nachr. Akad. Wiss. G\"ottingen, Math.
Phys. Kl IIa N.6, 87--120 (1955).}
\cita{Moser-1962}{J. Moser: {\it  On invariant curves of area-preserving
mappings of an annulus}, Nachr. Akad. Wiss. G\"ott,. II Math. Phys. Kl, 1--20 (1962).}
\cita{Moser-1967}{J. Moser: {\it  Convergent series expansions for
quasi-periodic motions}, Math. Ann., {\bf 169}, 136--176 (1967).}
\cita{Moser-1973}{J. Moser: {\it  Stable and random motions in dynamical
systems}, Princeton University press, Princeton (1973).}
\cita{Murray-1999}{N. Murray, M. Holman: {\it  The Origin of
Chaos in Outer Solar System}, Science, {\bf 283} Iss.5409,
1877 (1999).}
\cita{Nekhoroshev-1977}{N.N. Nekhoroshev: {\it  Exponential estimates of
the stability time of near-integrable Hamiltonian systems.} Russ. Math.
Surveys, {\bf 32}, 1 (1977).}
\cita{Nekhoroshev-1979}{N.N. Nekhoroshev: {\it Exponential estimates of
the stability time of near-integrable Hamiltonian systems, 2.} Trudy
Sem. Petrovs., {\bf 5}, 5 (1979).}
\cita{Robutel-1995}{P. Robutel: {\it  Stability of
the Planetary Three-Body Problem --- II. KAM Theory and Existence of
Quasiperiodic Motions}, Celestial Mechanics and Dynamical Astronomy,
{\bf 62}, 219--261 (1995).}
\cita{Standish-1990.1}{E.M. Standish Jr.: {\it  The observational basis
for JPL's DE 200, the planetary ephemerides of the Astronomical
Almanac}, Astronomy and Astrophysics, {\bf 233}, 252--271 (1990).}
\cita{Standish-1990.2}{E.M. Standish Jr.: {\it  An approximation to the
outer planet ephemeris error in JPL's DE 200}, Astronomy and
Astrophysics, {\bf 233}, 272--274 (1990).}

\def\testos{A. Giorgilli, U. Locatelli and M. Sansottera}
\def\testod{Kolmogorov and Nekhoroshev theory $\ldots$}

\title{Kolmogorov and Nekhoroshev theory for the problem of three bodies}

\author{\it ANTONIO GIORGILLI
\hfill\break Dipartimento di Matematica, Universit\`a degli Studi di Milano,
\hfill\break via Saldini 50, 20133\ ---\ Milano, Italy.}

\author{\it UGO LOCATELLI
\hfill\break Dipartimento di Matematica, Universit\`a degli Studi di Roma,
\hfill\break ``Tor Vergata'', via della Ricerca
Scientifica 1, 00133 Roma, Italy.}

\author{\it MARCO SANSOTTERA
\hfill\break Dipartimento di Matematica, Universit\`a degli Studi di Milano,
\hfill\break via Saldini 50, 20133\ ---\ Milano, Italy.}

\abstract{We investigate the long time stability in Nekhoroshev's
sense for the Sun--Jupiter--Saturn problem in the framework of the
problem of three bodies.  Using computer algebra in order to perform
huge perturbation expansions we show that the stability for a time
comparable with the age of the universe is actually reached, but with
some strong truncations on the perturbation expansion of the
Hamiltonian at some stage.  An improvement of such results is
currently under investigation.}

\section{1}{Introduction}
The stability of the Solar System is a classical, long standing and
challenging problem, already pointed out by Newton.  In this article we
revisit the problem in the light of the theorems of Kolmogorov and of
Nekhoroshev, with the aim of proving that they apply to the problem of
three bodies with the masses and orbital parameters of Jupiter and Saturn.

Let us briefly recall the historical development of our knowledge.
After 1954 a possible solution was suggested by the celebrated theorem
of Kolmogorov\bibref{Kolmogorov-1954} stating the existence of a large
measure set of invariant tori for a nearly integrable Hamiltonian
system, e.g., the planetary system when the mutual perturbation of the
planet is taken into account.  The relevance of Kolmogorov's result for
the planetary problem has been soon emphasized by
Arnold\bibref{Arnold-1963} and Moser\bibref{Moser-1973}.  In
particular Arnold worked out a proof taking into account the
degeneration of the unperturbed Hamiltonian which occurs in the
planetary case\bibref{Arnold-1963.1}.  On the other hand, Moser first
gave a proof for the case of an area preserving mapping of an
annulus\bibref{Moser-1962}, and a few years later pointed out that the
theorem of Kolmogorov implies that the classical Lindstedt series are
actually convergent\bibref{Moser-1967}.

As a matter of fact it was soon remarked by H\'enon that the
application of the Kolmogorov's theorem to the planetary motions is
not straightforward, due to the condition that the masses of the
planets should be small enough. Indeed, the available estimates could
only assure the applicability, e.g., to the problem of three bodies,
if the masses of the planets are less than that of a proton.  On the
other hand, numerical integrations of the full Solar System over a
time span of billions of years have shown that the orbits of the inner
planets exhibit a chaotic evolution which is incompatible with the
quasi periodic motion predicted by Kolmogorov's
theorem\bibref{Carpino-1987}\bibref{Laskar-1989}\bibref{Laskar-1994}.
Furthermore the subsystem of the major planets (i.e. Jupiter, Saturn,
Uranus and Neptune) shows a very small positive Lyapunov
exponent\bibref{Murray-1999}, once again, this cannot fit with a
motion on an invariant torus.

A second approach was suggested by Moser\bibref{Moser-1955} and
Littlewood\bibref{Littlewood-1959}\bibref{Littlewood-1959.1} and fully
stated by
Nekhoroshev\bibref{Nekhoroshev-1977}\bibref{Nekhoroshev-1979} with his
celebrated theorem on exponential stability.  According to this
theorem the time evolution of the actions of the system (which in the
planetary case are actually related to the semimajor axes, the
inclinations and the eccentricities) remains bounded for a time
exponentially increasing with the inverse of the perturbation
parameter.  Thus, although the possibility of a chaotic motion is not
excluded, nevertheless a dramatic change of the orbits should not
occur for such a long time, and it may be conjectured that such a time
exceeds the age of the Solar System itself.  But also in this case the
problem of the applicability of the theorem still persists, since the
analytical estimates based on Nekhoroshev's formulation or other
analytical proofs give ridiculous estimates for the size of the masses
of the planets.

In recent years the estimates for the applicability of both
Kolmogorov's and Nekhoroshev's theorems to realistic models of some
part of the Solar System have been improved by some authors.  For
example the applicability of Nekhoroshev's theorem to the stability of
the Trojan asteroids in the vicinity of the triangular Lagrangian
points has been investigated by Giorgilli et
al.\bibref{Giorgilli-1989}\bibref{Celletti-1991}\bibref{Giorgilli-1997}),
Efthimiopoulos~et~al.\bibref{Efthimiopoulos-2005} and by
Lhotka~et~al.\bibref{Lhotka-2008}, the connection between
Nekhoroshev's theorem and Arnold diffusion has been considered by
Efthimiopoulos\bibref{Efthimiopoulos-2008}; the applicability of
KAM theorem has been studied by Robutel\bibref{Robutel-1995},
Fejoz\bibref{Fejoz-2005}, Celletti~et~al.\bibref{Celletti-2005},
Gabern~et~al.\bibref{Gabern-2005} and by Locatelli and
Giorgilli\bibref{Loc-Gio-2005}\bibref{Locatelli-2007}.  In
particular in the latter two articles the Sun--Jupiter--Saturn
(hereafter SJS) system is investigated, and evidence is produced that
an invariant torus exists in the {\it vicinity of the initial data}
of Jupiter and Saturn, at least in the approximation of the general
problem of three bodies.

In the present article we study the stability in Nekhoroshev's sense of
the neighbourhood of the invariant torus for the SJS system.  The aim
is to give evidence, with help of a computer-assisted calculation,
that the size of the neighbourhood of the invariant torus for which
exponential stability holds for a time interval as long as the age of
the universe is big enough to contain the {\it actual initial data} of
Jupiter and Saturn.  We should say that such an ambitious goal is
still out of our actual possibilities.  However, we show that our
methods should allow us to achieve our program provided a sufficient
computer power will be available in the next future and a further
refinement of our approximation methods will be worked out.  This is
work for the next future.

\section{2}{Theoretical framework}
The basis of our approach is the investigation of the stability of a
neighbourhood of an invariant Kolmogorov's torus.  To this end let us
briefly recall the statement of Kolmogorov's theorem

\theorem{kam}{Consider a canonical system with Hamiltonian
$$
H(p,q) = h(p) + \epsilon f(p,q)\ .
\formula{1}
$$
Let us assume that the unperturbed part of the Hamiltonian is
non-degenerate, i.e., $\det\left(\frac{\partial^2 h}{\partial
p_j\partial p_k}\right)\ne 0\,$, and that $p^*\in\reali^n$ is such
that the corresponding frequencies $\omega=\frac{\partial h}{\partial
p}(p^*)$ satisfy a Diophantine condition, i.e.,
$$
\bigl|\langle k,\omega\rangle\bigr| \ge \gamma |k|^{-\tau}
\quad \forall\ 0\ne k\in\interi^n
$$
with some constants $\gamma\gt 0$ and $\tau\ge n-1$\ .  Then for
$\epsilon$ small enough the Hamiltonian~\frmref{1} admits an
invariant torus carrying quasiperiodic motions with frequencies
$\omega$.  The invariant torus lies in a $\epsilon$-neighbourhood of
the unperturbed torus $\{(p,q)\,:\> p=p^*,\>q\in\toro^n\}$.}\endclaim

The question is about the dynamics in the neighbourhood of the
invariant torus.  In order to discuss this point we need a few
technical details about the Kolmogorov's proof method.  The key points,
clearly outlined in the original short note~\dbiref{Kolmogorov-1954},
are the following.  First, one picks an unperturbed invariant torus $p^*$
for the Hamiltonian~\frmref{1} characterized by diophantine
frequencies $\omega$, and expands the Hamiltonian in power series
of the actions $p$ in the neighbourhood of $p^*$.  Thus (with a
translation moving $p^*$ to the origin of the actions space) one gives
the initial Hamiltonian the form

$$
H(p,q) =  \langle\omega,p\rangle + \epsilon A(q) 
+ \epsilon\bigl\langle B(q),p\bigr\rangle
+ \frac{1}{2} \bigl\langle Cp,p\bigr\rangle + O(p^2)
\formula{2}
$$
where $C=\left[\frac{\partial^2 h}{\partial p_j\partial
p_k}(p^*)\right]$ is a symmetric matrix, and $A(q)$ and $\bigl\langle
B(q),p\bigr\rangle$ are the terms independent of $p$ and linear in $p$
in the power expansion of the perturbation $f(p,q)\,$,
respectively. The quadratic part in $\Oscr(p^2)$ is of order $\epsilon$,
too.  The next step consists in performing a near the identity
canonical transformation which gives the Hamiltonian the Kolmogorov's
normal form
$$
H'(p',q') = \langle\omega,p'\rangle + O({p'}^2)\ .
\formula{3}
$$
As Kolmogorov points out, the invariance of the torus $p'=0$ is
evident, due to the particular form of the normalized Hamiltonian.
The whole process requires a composition of an infinite sequence of
transformations, and the most difficult part is to prove the
convergence of such a sequence.  The point which is of interest to us
is that {\it the transformed Hamiltonian~\frmref{3} is analytic in a
neighbourhood of the invariant torus $p'=0\,$.}

Let us emphasize that the analytical form of the Hamiltonian~\frmref{3}
is quite similar to that of a Hamiltonian in the neighbourhood of an
elliptic equilibrium, namely
$$
H(x,y) = \frac{1}{2} \sum_{j=1}^{n} \omega_j \bigl(x_j^2+y_j^2\bigr) +
\ldots\ ,
$$
where the dots stand for terms of degree larger than 2 in the Taylor
expansion.  For, introducing the action-angle variables $p,q$ via the
usual canonical transformation $x_j =\sqrt{2p_j}\cos q_j\,$, $y_j
=\sqrt{2p_j}\sin q_j\,$, the latter Hamiltonian takes essentially the
form~\frmref{3}.  Thus the exponential stability of the invariant
torus $p'=0$ may be proved using the theoretical scheme that works fine
in the case of an elliptic equilibrium, e.g., in the case the
triangular Lagrangian points.

As a matter of fact, a much stronger result holds true, namely that
the invariant torus is superexponentially stable, as stated
in~\dbiref{Morbidelli-1995} and~\dbiref{Giorgilli-1997.5}.  However, a
computer assisted method for the theory of superexponential stability
seems not be currently available, so we limit our study to the
exponential stability in Nekhoroshev's sense.

\section{3}{Technical tools}
Let us now come to the improvement of the estimates for the
applicability of the theorems of Kolmorogov and Nekhoroshev.  The key
point is to use an explicit construction of the normal form up to a
finite order with algebraic manipulation in order to reduce the size
of the perturbation, and then apply a suitable formulation of the
theorems.

Let us explain this point by making reference to the theorem of
Kolmogorov.  Starting with the Hamiltonian~\frmref{2}, we perform a
finite number, $r$ say, of normalization steps in order to give the
Hamiltonian the normal form up to order $r$
$$
H^{(r)}(p,q) = \langle\omega,p\rangle + \frac{1}{2}\langle C p,p\rangle
 + \epsilon^r A^{(r)}(q)
  + \epsilon^r\langle B^{(r)}(q),p\rangle + {\cal R}^{(r)}(p,q)
\formula{10}
$$ 
with ${\cal R}^{(r)}(p,q) = \Oscr(|p|^2)$, so that {\it the perturbation
is now of order $\epsilon^r$.}  

To this end we implement the normalization algorithm for the normal
form of Kolmogorov step by step in powers of $\epsilon$, as in the
traditional expansions in Celestial Mechanics.  The full justification
of such a procedure, including the convergence proof, is given, e.g.,
in~\dbiref{Giorgilli-1997.1} and~\dbiref{Gio-Loc-1997.1}.  The
resulting Hamiltonian has still the form~\frmref{2}, with, however,
$\epsilon$ replaced by $\epsilon^r$.  Thus, a straightforward
application of the theorem reads, in rough terms: {\it if
$\epsilon^r\lt\epsilon_*$, then an invariant torus exists.}  The power
$r$ may considerably improve the estimate of the threshold for the
applicability of the theorem.  This approach has been translated in a
computer assisted rigorous proof, which has been successfully applied
to a few simple
models\bibref{Celletti-2000}\bibref{Loc-Gio-2000}\bibref{Gabern-2005}).

Let us now come to the part concerning the estimate of the stability
time which is the main contribution of the present note.  To this end
we remove from the Hamiltonian~\frmref{10} all the contributions which
are independent of or linear in the actions $p$, namely the terms
$\epsilon^r A^{(r)}(q) +\epsilon^r\langle B^{(r)}(q),p\rangle$, which
are small, thus obtaining a reduced Hamiltonian in Kolmogorov's normal
form.  Moreover, we expand the perturbation ${\cal R}^{(r)}(p,q)$ in
power series of $p$ and Fourier series of $q$, thus getting a
Hamiltonian in the form
$$
H(p,q) = \langle\omega,p\rangle + H_1(p,q) + H_2 (p,q)+ \ldots\ .
\formula{4}
$$
where $H_s(p,q)$ is a homogeneous polynomial of degree $s+1$ in the
actions $p$ and a trigonometric series in the angles $q$.  Here, the
upper index $r$ of $H$ has been removed because it is now meaningless,
since we use the latter Hamiltonian as an approximation of
the Kolmogorov's normal form.

On this Hamiltonian we perform a Birkhoff normalization up to a finite
order, that we denote again by $r$ although it has no relation with
the order of Kolmogorov's normalization used above.  Thus we get a
Birkhoff normalized Hamiltonian
$$
H = \langle\omega,p\rangle + Z_1(p) + \ldots + Z_r(p) + \Fscr_r(p,q),
$$
with $\Fscr_r(p,q)$ a power series in $p$ starting with terms of degree
$r+2$.  We omit the details about this part of the calculation, since
there are a number of well known formal algorithms that do the job.
We concentrate instead on the quantitative estimates.

Let us introduce a norm for a function
$f(p,q)=\sum_{|l|=s,k\in\interi^n}f_{l,k}p^l e^{i\langle k,q \rangle}$
which is a homogeneous polynomial of degree $s$ in the actions $p$.
Precisely define
$$
\|f\| = \sum_{|l|=s\,,\,k\in\interi^n} |f_{l,k}|\ .
\formula{5}
$$
Moreover consider the domain 
$$
\Delta_\rho = \left\{ p\in\reali^n,\ |p_j|\leq\rho
\>,\>j=1\,,\,\ldots\,,\,n \right\}\ .
\formula{6}
$$

\noindent
Then we have
$$
|f(p,q)| \leq \|f\| \rho^s\quad {\rm for\  }p\in\Delta_\rho\>,\
 q\in\toro^n\ .
$$
Let now $p(0)\in\Delta_{\rho_0}$ with $\rho_0<\rho$.  Then we have
$p(t)\in\Delta_{\rho}$ for $|t|<T$, where $T$ is the escape time from
the domain $\Delta_\rho$.  This is the quantity that we want to
evaluate.  To this end we use the elementary estimate
$$
\left| p(t)-p(0) \right| \leq 
|t|\cdot\sup_{|p|<\rho}|\dot p| < |t|\cdot\bigl\| \{ p,\Fscr
 \}  \bigr\| {\rho^{r+2}}\ .
\formula{7}
$$
The latter formula allows us to find a lower bound for the escape time
from the domain $\Delta_\rho$, namely
$$
\tau(\rho_0,\rho,r) = 
 \frac{\rho-\rho_0}{\bigl\|\{ p,\Fscr \}\bigr\|{\rho^{r+2}}}
\ ,
\formula{11}
$$
which however depends on $\rho_0$, $\rho$ and $r$.  We emphasize that
in a practical application, e.g., to the SJS system, $\rho_0$ is fixed
by the initial data, while $\rho$ and $r$ are left arbitrary.  Thus we
try to find an estimate of the escape time $T(\rho_0)$ depending only
on the physical parameter $\rho_0$.  To this end we optimize
$\tau(\rho_0,\rho,r)$ with respect to $\rho$ and $r$, proceeding as
follows.  First we keep $r$ fixed, and remark that the function
$\tau(\rho_0,\rho,r)$ has a maximum for
$$
\rho=\frac{r+2}{r+1} \rho_0\ .
$$
This gives an optimal value of $\rho$ as a function of $\rho_0$ and
$r$, and so a new function 
$$
\tilde\tau(\rho_0,r) = \sup_{\rho\ge\rho_0} \tau(\rho_0,\rho,r)
$$ 
which is actually computed by putting the optimal value
$\rho=\rho_0(r+2)/(r+1)$ in the expression above for
$\tau(\rho_0,\rho,r)$.  Next we look for the optimal value $r_{\rm
opt}$ of $r$, which maximizes $\tilde\tau(\rho_0,r)$ when $r$ is
allowed to change.  That is, we look for the quantity
$$
T(\rho_0) = \max_{r\ge 1} \tilde\tau(\rho_0,r)\ ,
$$
which is our best estimate of the escape time, depending only on the
initial data.  We define the latter quantity as the estimated
stability time.  Here we need some further considerations in order to
convince the reader that the maximum in the r.h.s.~of the latter
formula actually exists.  This follows from the asymptotic properties
of the Birkhoff's normal form.  For, according to the available
analytical estimates based on Diophantine inequalities for the
frequencies, the norm $\bigl\|\{ p,\Fscr_r\}\bigr\|$ in the denominator
of~\frmref{11} is expected to grow as $(r!)^n$, $n$ being the number
of degrees of freedom.  Thus, for $\rho_0$ small enough the
denominator $\bigl\|\{ p,\Fscr_r\}\bigr\|{\rho_0^r}$ reaches a minimum
for some $r^n\sim 1/\rho_0$, which means that the wanted maximum
actually exists, thus providing the optimal value $r_{\rm opt}\,$.  We
also remark that although no proof exists that the analytical
estimates are optimal, accurate numerical investigations based on
explicit expansions show that the $r!$ growth of the norms actually
shows up (see, e.g.,~\dbiref{Contopolus-2003}
and~\dbiref{Contopolus-2004}).  Working out an analytical evaluation
of the stability time on the basis of these considerations leads to an
exponential estimate of Nekhoroshev's type for $T(\rho_0)$ (see,
e.g.,~\dbiref{Giorgilli-1989}). Here we replace the analytical
estimates with an explicit numerical optimization of
$\tilde\tau(\rho_0,r)$ by just calculating it for increasing values of
$r$ until the maximum is reached.

Our aim is to perform the procedure above by using computer algebra.
Thus some truncation of the functions must be introduced in order to
implement the actual calculation.  The most straightforward approach
is the following.  First we truncate the Hamiltonian~\frmref{4} at a
finite polynomial order in the actions.  This is legitimate if the
radius $\rho$ of the domain is small, due to the well known properties
of Taylor series.  However, the Fourier expansion of every term $H_s$
still contains infinitely many contributions.  Here we take advantage
of the exponential decay of the Fourier coefficients of analytic
functions and of some algebraic properties of the Poisson brackets.
Precisely, let $f(q)=\sum_{k} f_k e^{i\langle k,q\rangle}$; the
dependence of the coefficients $f_k$ on the actions is unrelevant
here.  Then the exponential decay of the coefficients means that
$|f_k| \le Ce^{-|k|\sigma}$ with some positive constants $C$ and
$\sigma$.  Thus, having fixed a positive integer $K$ we truncate the
Fourier expansion as $f(q)=\sum_{|k|\le K} f_k e^{i\langle
k,q\rangle}$, i.e., we remove all Fourier modes $|k|\gt K$.  This is
allowed because the exponential decay assures that the neglected part
is small.  The interested reader may find a more detailed discussion
about this method of splitting the Hamiltonian
in~\dbiref{Giorgilli-03.1}.

Coming back to our problem, we include in $H_s(p,q)$ all Fourier
coefficients with $|k|\le sK$, so that $H_s(p,q)$ is a homogeneous
polynomial of degree $s+1$ in the actions $p$ and a trigonometric
polynomial of degree $sK$ in the angles $q$.  The algebraic property
mentioned above is that such a splitting of the Hamiltonian is
preserved by the Lie series algorithm that we apply through all our
calculations.  This in view of the elementary fact that the Poisson
bracket between two functions, $f_r$ and $f_s$ say, which are
homogeneous polynomial of degree $r+1$ and $s+1$, respectively, in $p$
and trigonometric polynomials of degree $rK$ and $sK$, respectively,
in $q$ produces a new function of degree $r+s+1$ in $p$ and $(r+s)K$
in $q$.

A final remark concerns the estimate of the remainder $\Fscr_r$
in~\frmref{7}, which is an infinite series, too.  Here we just
calculate the first term of the remainder, namely the term of degree
$r+1$, and multiply its norm by a factor $2$.  This factor is
justified in view of the fact that the analytical estimates of the
same quantities involve a sum of a geometric series which, for $\rho$
small enough, decreases with a ratio less than $1/2$.  Here a natural
objection could be that for some strange reason the norm of the
remainder at some finite order could be smaller than predicted
by the analytical estimates.  However, it is a common experience that
after a few perturbation steps the norms of the functions take a
rather regular behaviour consistent with the geometric decrease
predicted by the theory.  Thus, our choice appears to be justified by
experience.

As a final remark we note that our way of dealing with the truncation
is the most straightforward one, but it is not the sole possible.
Other more refined criteria may be invented, of course, which may take
into account the most important contribution while substantially
reducing the number of coefficients to be calculated.  In this sense
our direct approach should be considered as a first attempt to check
if the concept of Nekhoroshev's stability may be expected to apply to
our Solar System.  Although being unable to produce rigorous results
in a strict mathematical sense, we believe that our method gives
interesting results in the spirit of classical perturbation methods.

\section{4}{Application to the planetary problem of three bodies}
Applying the theories of Kolmokorov and Nekhoroshev to the planetary
problem is not straightforward, due to the degeneration of the
Keplerian motion.  In order to remove such a degeneration, a lenghty
procedure is needed; this essentially requires a suitable adaptation
of the canonical coordinates, paying a very particular care to the
secular ones (to appreciate some deep point of view about this
problem, see, e.g.,~\dbiref{Arnold-1963.1}
and~\dbiref{Nekhoroshev-1977}).

In our approach the difficulty shows up in the part concerning the
application of Kolmogorov's theory.  Once a Kolmogorov torus has been
constructed, then there is no extra difficulty in applying the method
of sect.~\secref{3}, due to the fact that the method is local.  In the
present section we give a brief sketch of the procedure for the
construction of a Kolmogorov torus.  The complete procedure is
described in~\dbiref{Loc-Gio-2005} and~\dbiref{Locatelli-2007}, to
which we refer for details.

Following a traditional approach, we first reduce the integrals of
motion (i.e. the linear and angular momenta); therefore, we separate
the fast variables (essentially the semimajor axes and the mean
anomalies) from the slow ones (the eccentricities and the inclinations
with the conjugated longitudes of the perihelia and of the nodes).
This is usually done in Poincar\'e variables by writing a reduced
Hamiltonian of the form 
$$
H^{R}(\Lambda,\lambda,\csi,\eta)= F^{(0)}(\Lambda)+
\mu F^{(1)}(\Lambda,\lambda,\csi,\eta)
\formula{Ham-3cp-iniz}
$$
with
$$
\mu=\max\{m_1\,/\,m_0\,,\,m_2\,/\,m_0\,\}
$$
where $m_0$ is the mass of the  star, $m_1$ and $m_2$ are
the masses of the planets, $\Lambda_j,\,\lambda_j\,$ are the fast
variables and $\xi_j,\,\eta_j\,$ are the slow (Cartesian-like)
variables. Here, obviously, the values of the index $j=1\,,\,2$
correspond to the internal planet and to the external one,
respectively. 

On this Hamiltonian we perform a procedure which is the natural
extension of the one devised by Lagrange and Laplace in order to
calculate the secular motion of the eccentricities and the
inclinations and the conjugated angles.

The first step is the identification of a good unperturbed invariant
torus for the fast angles $\lambda$, setting for a moment the slow
variables $\xi,\eta$ to zero.  Here is a short description.

\item{(i)}Having fixed a frequency vector $n^*\in\reali^2$, 
we determine the corresponding action values $\Lambda^*$ corresponding
to a torus which is invariant for an integrable approximation of the
system, where the dependency on both the fast angles $\lambda$ and on
the secular coordinates $\xi,\eta$ is dropped. This can be done by
solving the equation
$$
\frac{\partial\, \langle H^{R}\rangle_{\lambda}}{\partial \Lambda_j}
\Bigg|_{{\scriptstyle \Lambda=\Lambda^*}
\atop{\scriptstyle \csi\,,\,\eta=0}}=n_j^*\ ,
\quad j=1,\, 2\,.
$$
Here $\langle H^{R}\rangle_{\lambda} =
\frac{1}{4\pi^2}\int_{\toro^2} H^{R}\,d\lambda_1\,d\lambda_2 $ is
the average of the Hamiltonian $H^{R}$ with respect to the fast
angles. The explicit value of $n^*$ is chosen so that it reflects
the true mean motion frequencies of the planets (see next section for
our values).  Having solved the previous equation with respect to the
unknown vector $\Lambda^*$, we expand $H^{R}$ in power series of
$\Lambda-\Lambda^*$.  With a little abuse of notation we denote again
by $\Lambda$ the new variables.

\item{(ii)}We perform two further canonical transformations which
make the torus 
$\Lambda=\xi=\eta=0$ to be invariant up to order 2 in the
masses.  Indeed, these changes of coordinates are borrowed from the
Kolmogorov's normalization algorithm, but we look for a Kolmogorov's
normal form with respect to the fast variables only, considering the
slow ones essentially as parameters, although they are changed too.
More precisely, we determine  generating functions of the form 
$\chi_{j}(\Lambda,\lambda,\xi,\eta)=\Lambda^{j-1}g_j(\lambda,\xi,\eta)$
for $j=1,2\,$, where $g_j(\lambda,\xi,\eta)$ includes a finite order
expansion both in Fourier modes with respect to the fast angles
$\lambda$ and in polynomial terms of the slow variables
$\xi,\eta\,$.  The aim of this step is to reduce the size of terms
independent of or linear in the fast actions so that it is of the same
order as the rest of the perturbation.  We denote by  $H^T$ the
resulting Hamiltonian, which is still trigonometric in the fast angles
$\lambda$ and polynomial in  $\Lambda,\xi,\eta$.

\noindent
The next goal is to determine a good invariant torus for the slow
variables $\xi,\eta\,$.  To this end we combine the classical Lagrange's
calculation of the secular frequencies with a Birkhoff's procedure
that takes into account the nonlinearity.

\item{(iii)}We consider the secular system, namely the average
$\langle H^T\rangle$ of the Hamiltonian $H^{T}$ resulting from the
step (ii) above.  Acting only on the quadratic part of the Taylor
expansion of $\langle H^T\rangle$ in $\xi,\eta$ we determine a first
approximation of the secular frequencies, and transform the
Hamiltonian so that its quadratic part has a diagonal form.  This part
of the calculation follows the lines of Lagrange's theory, but the
calculation is worked out at the second order approximation in the
masses.  The diagonalization of the quadratic part requires a linear
canonical transformation, which is a standard matter.  Thus the
quadratic part in $\xi,\eta$ of the resulting Hamiltonian has the form
$\frac{1}{2}\sum_{j}\nu_j (\xi_j^2 +\eta_j^2)$, where $\nu$ are the
secular frequencies and we denote again by $\xi,\eta$ the slow
variables.

\item{(iv)}We perform a Birkhoff's normalization 
up to order 6 in $\xi,\eta$.  This gives a normalized secular
Hamiltonian $H^{B}$ which in action-angle variables
$\xi_j=\sqrt{2I_j}\cos\phi_j\>,\eta_j=\sqrt{2I_j}\sin\phi_j$ takes
the form
$$
H^B = \nu\cdot I + h^{(4)}(I)+h^{(6)}(I) + F(\Lambda,I,\phi)\ ,
$$
where $h^{(4)}$ and $h^{(6)}$ are polynomials of degree 2 and 3 in
$I$, respectively.  This step removes the degeneration of the secular
motion, thus allowing us to take into account the nonlinearity of the
secular part of the problem.

\item{(v)}Having fixed the slow frequencies $g^*$ so that they
reflect the true frequencies of the system, we determine a secular
torus $I^*$ corresponding to these frequencies, by using the
integrable approximation of $H^B$.  This is done by solving for $I$
the equation
$$
\frac{\partial\, h^{(4)}}{\partial I_j}(I)+
\frac{\partial\, h^{(6)}}{\partial I_j}(I)=
g_j^*-\nu_j^*\ ,\qquad j=1,\,2\ .
$$

\noindent
The values $\Lambda^*$ and $I^*$ so determined provide the first
approximation of the Kolmogorov's invariant torus.  Reintroducing the
fast angles and performing on the original Hamiltonian $H^R$ all the
transformations that we have done throughout our procedure (i)--(v) we
get a Hamiltonian of the form~\frmref{2} which is the starting point
for Kolmogorov's normalization algorithm.  After a number of
Kolmogorov's steps the Hamiltonian takes the form~\frmref{10}, thus
giving a good approximation of an invariant torus with frequencies
$n^*$ and $g^*$.  The latter form is precisely the output of the
calculation illustrated in~\dbiref{Gio-Loc-1997.1}, and by removing
all terms which are independent of or linear in the actions $p$ it
provides a Hamiltonian as that in~\frmref{4}.  This is the starting
point for our algorithm evaluating the stability time in the
neighbourhood of the invariant torus.

\table{tab1}{Physical parameters for the Sun--Jupiter--Saturn system
taken from JPL at the Julian Date $2451220.5\,$.}
{
\vbox{\tabskip=0pt\offinterlineskip
\def\tablerule{\noalign{\hrule}}
\halign{
\vrule\hfil$\>${#}\ {}
&\strut{#}$\>$\hfil\vrule
&$\>$\hfil{#}$\>$\hfil\vrule
&$\>$\hfil{#}\hfil$\>$\vrule\ignorespaces
\cr
\tablerule
\phantom{\vbox to 12pt{\relax}}
  & & Jupiter ($j=1$) & Saturn ($j=2$)\cr
\tablerule
\phantom{\vbox to 14pt{\relax}}mass
 & $m_j$ & $(2\pi)^2/1047.355$ & $(2\pi)^2/3498.5$  \cr
\tablerule
\phantom{\vbox to 12pt{\relax}}semi-major axis
  & $a_j$ & $5.20092253448245$ & $9.55716977296997$ \cr
\tablerule
\phantom{\vbox to 12pt{\relax}}mean anomaly
  & $M_j$ & $6.14053316064644$ & $5.37386251998842$ \cr
\tablerule
\phantom{\vbox to 12pt{\relax}}eccentricity
  & $e_j$ & $0.04814707261917873$ & $0.05381979488308911$  \cr
\tablerule
\phantom{\vbox to 12pt{\relax}}perihelion argument
  & $\omega_j$ & $1.18977636117073$ & $5.65165124779163$  \cr
\tablerule
\phantom{\vbox to 12pt{\relax}}inclination
  & $i_j$ & $0.006301433258242599$ & $0.01552738031933247$  \cr
\tablerule
\phantom{\vbox to 12pt{\relax}}longitude of the node
  & $\Omega_j$ & $3.51164756250381$ & $0.370054908914043$  \cr
\tablerule
}}
}

\table{tab2}{The frequencies of the unperturbed torus in the SJS
system corresponding to the initial data and physical parameters in
table~\tabref{tab1}. The values are calculated via frequency analysis
on the orbits obtained by direct integration of the equations for the
problem of three bodies.}{
\vbox{\tabskip=0pt\offinterlineskip
\def\tablerule{\noalign{\hrule}}
\halign{
\vrule\hfil$\>${#}$\>$\vrule
&$\>$\hfil\strut{#}$\>$\hfil\vrule
&$\>$\hfil{#}\hfil$\>$\vrule\ignorespaces
\cr
\tablerule
\phantom{\vbox to 14pt{\relax}}
  & Jupiter & Saturn\cr
\tablerule
\phantom{\vbox to 14pt{\relax}}fast frequencies
& $n_1^*=0.52989041594442$
& $n_2^*=0.21345444291052$  \cr
\tablerule
\phantom{\vbox to 14pt{\relax}}secular frequencies
& $\phantom{\mu}g_1^*=-0.00014577520419$
& $\phantom{\mu}g_2^*=-0.00026201915143$ \cr
\tablerule
}}
}

\section{5}{Application to the Sun--Jupiter--Saturn system}
We come now to the application of our procedure to the SJS system.
Let us first define the model.  We consider the general problem of
three bodies with the Newtonian potential.  Thus, the contribution due
to the other planets of the Solar System is not taken into account in
our approximation.  The expansion of the Hamiltonian is a classical
matter, so we skip the details, just recalling that all the expansions
have been done via algebraic manipulation, using a package developed
on purpose by the authors.

The choice of the model plays a crucial role in determining the
frequencies of the torus, that we calculate by integrating the Newton
equations for the problem of three bodies and applying the frequency
analysis (see, e.g.,~\dbiref{Laskar-99}) to the computed orbit.  As
initial data we take the orbital elements of Jupiter and Saturn as
given by JPL\footnote{1}{The data about the planetary motions
provided by the Jet Propulsion Laboratory are publicly available
starting from the webpage {\tt http://www.jpl.nasa.gov/}} 
for the
Julian Date $2451220.5\,$.  This is the point where the connection
with the physical parameters of our Solar System is made.  The
physical parameters and the orbital elements are reported in
table~\tabref{tab1}.  The calculated frequencies are given in
table~\tabref{tab2}.

The choice of the Julian Date $2451220.5\,$ in order to set the
initial data is completely arbitrary, of course, its sole
justification being that such data are directly available from JPL.
Choosing different dates or different determinations of the planet's
elements could lead to a slightly different determination of the
frequencies, and so also of the invariant torus.  However, we
emphasize that the aim of the present work is precisely to give a long
time stability result which applies to a neighbourhood of the
invariant torus.  The size of such a neighbourhood should be large
enough to cover the unavoidable uncertainty in determining the initial
data for the SJS system.  This is a delicate matter, of course,
because the JPL data reflect the dynamics of the full Solar System,
while our study is concerned only with the model of three bodies.
However, we may get some hint on the size of the uncertainty precisely
by looking at the JPL data.

\table{tab3}{Estimates of the uncertainties on the initial values 
of the canonical coordinates $(\Lambda,\lambda,\csi,\eta)\,$. These
evaluations are derived from the comparison of different sets of JPL's
DE.}{
\vbox{\tabskip=0pt\offinterlineskip
\def\tablerule{\noalign{\hrule}}
\halign{
\vrule\hfil$\>${#}$\>$\vrule
&$\ $\hfil\strut{#}$\>$\hfil\vrule
&$\ $\hfil\strut{#}$\>$\hfil\vrule
&$\ $\hfil\strut{#}$\>$\hfil\vrule
&$\ $\hfil\strut{#}\hfil$\>$\vrule\ignorespaces
\cr
\tablerule
\phantom{\vbox to 14pt{\relax}}
  & $\Delta\Lambda_j$ & $\Delta\lambda_j$
  & $\Delta\csi_j$ & $\Delta\eta_j$ \cr
\tablerule
\phantom{\vbox to 14pt{\relax}}Jupiter ($j=1$)
& $1.8\,\times 10^{-6}$
& $6.6\,\times 10^{-5}$
& $1.1\,\times 10^{-5}$
& $2.8\,\times 10^{-6}$  \cr
\tablerule
\phantom{\vbox to 14pt{\relax}}Saturn ($j=2$)
& $1.7\,\times 10^{-6}$
& $3.0\,\times 10^{-5}$
& $3.3\,\times 10^{-6}$
& $3.2\,\times 10^{-6}$ \cr
\tablerule
}}
}

As everybody knows, the initial positions and velocities of the
planets are usually taken from the Development Ephemeris of the Jet
Propulsion Laboratory (for short, JPL's DE). There are several sets of
these ephemerides, each version of them being based on more and more
observational data, which take benefit from the improvement of the
techniques.  Thus, each new version of the JPL's DE is expected to
improve the precision of the data with respect to the older ones and,
then, one can approximately evaluate the error of the older versions
by comparison with the most recent one~(see the initial discussion
in~\bibref{Standish-1990.2}).  The positions and velocities of the
planets given by five different sets of JPL's DE are listed in
table~15 of Standish's paper~\dbiref{Standish-1990.1}.  For each kind
of these data we can determine a narrow interval containing all of
them and we can calculate the Keplerian orbital elements corresponding
to the extrema of such intervals.  By applying all the necessary
transformations we translate these data into uncertainties for the
Poincar\'e's canonical coordinates $(\Lambda,\lambda,\xi,\eta)$ that
have been used in order to write the
Hamiltonian~\frmref{Ham-3cp-iniz}. These uncertainties are reported in
table~\tabref{tab3}.  This provides us with a first approximation of the
neighbourhood of our initial data that contains all JPL's DE reported
in Standish's paper.  We should now apply all the canonical
transformations needed in order to construct an invariant torus close
to the SJS orbit.  However we remark that all such transformations are
very smooth, being analytic, volume preserving, and most of them are
close to identity, so that they add just a small correction with
respect to the data in table~\tabref{tab3}.  Thus we may confidently
expect that at some time the phase space point representing the
position of the SJS system lies in a neighbourhood of our approximated
invariant torus the size of which is evaluated to be
$\Oscr(10^{-6})\,$ for the fast actions and $\Oscr(10^{-5})\,$ for the
secular coordinates.

\table{tab4}{Maximal discrepancies about the orbital
  elements of the SJS system between a numerical integration and the
  semi-analytic one, that is based on the construction of the
  invariant torus corresponding to the frequencies values given in
  table~\tabref{tab2}. The maximal relative errors on the semi-major
  axis $a_j$ and on the eccentricities $e_j$ are reported here for
  both Jupiter (corresponding to $j=1$) and Saturn (i.e., $j=2$);
  the same is made also for the maximal
  absolute errors on the ``fast angle'' $\lambda_j=M_j+\omega_j$ and on
  the perihelion argument $\omega_j\,$. In the present case, the
  comparisons are made starting from the initial conditions given in
  table~\tabref{tab1} and for a time span of 100 Myr.}{
\vbox{\tabskip=0pt\offinterlineskip
\def\tablerule{\noalign{\hrule}}
\halign{
\vrule\hfil$\>${#}$\>$\vrule
&$\ $\hfil\strut{#}$\>$\hfil\vrule
&$\ $\hfil\strut{#}$\>$\hfil\vrule
&$\ $\hfil\strut{#}$\>$\hfil\vrule
&$\ $\hfil\strut{#}\hfil$\>$\vrule\ignorespaces
\cr
\tablerule
\phantom{\vbox to 14pt{\relax}}
& ${\rm Max}_t \left\{\left|\frac{\Delta a_j(t)}{a_j(t)}\right|\right\}$
& ${\rm Max}_t \left\{\left|\Delta \lambda_j(t)\right|\right\}$
& ${\rm Max}_t \left\{\left|\frac{\Delta e_j(t)}{e_j(t)}\right|\right\}$
& ${\rm Max}_t \left\{\left|\Delta \omega_j(t)\right|\right\}$\cr
\tablerule
\phantom{\vbox to 14pt{\relax}}Jupiter
& $1.5\,\times 10^{-6}$
& $5.0\,\times 10^{-4}$
& $1.3\,\times 10^{-3}$
& $1.3\,\times 10^{-3}$\cr
\tablerule
\phantom{\vbox to 14pt{\relax}}Saturn
& $6.8\,\times 10^{-6}$
& $1.1\,\times 10^{-3}$
& $4.3\,\times 10^{-3}$
& $7.3\,\times 10^{-3}$\cr
\tablerule
}}
}

Let us now come back to the actual calculation.  The Kolmogorov's
normal form has been computed up to order $17$, with the generating
function exhibiting a good geometric decay.  Furthermore, we have
compared the orbit on the approximate invariant torus with that
produced by a direct numerical integration of the equations of motion,
thus finding a quite good agreement between them, as shown in
table~\tabref{tab4}.  Here, we omit the details about these lengthy
calculations, since a complete report has been already given
in~\dbiref{Locatelli-2007}.

The calculation of Kolmogorov's normal form produces a Hamiltonian
which is analytic in the neighbourhood of the approximated invariant
torus.  Our program performs the calculation of this Hamiltonian with
the polynomial series in the actions truncated at order $3$ and the
trigonometric series truncated at order $34$
(see~\dbiref{Locatelli-2007} for more details).  On this Hamiltonian
we would like to apply the procedure of sect.~\secref{3}.  However, a
major obstacle raises up: the number of coefficients in the series
that we have calculated is more than $7\,100\,000$.  Such a huge
number of coefficients can not be handled in a Birkhoff normalization
procedure.  For, referring to the discussion at the end of
sect~\secref{3} we should set the parameter $K$ for the truncation of
trigonometric series to $34$, thus getting a truncation at
trigonometric degree $68,\,102,\,\ldots,\,34 r,\,\ldots$ at successive
order.  A rough estimate of the number of generated coefficients shows
that we shall soon run out of memory and of time on any available
computer.  Thus, we must introduce some further approximation.

In view of the considerations above we decided, as a first approach,
to strictly follow the truncation scheme illustrated at the end of
sect.~\secref{3} by just lowering the value of $K$.  We report the
results of this first attempt, which in our opinion appear already to
be interesting.  Thus we expand the Hamiltonian in the form
$$
H(p,q) = \langle\omega,p\rangle + H_1(p,q) + H_2 (p,q)
$$
by keeping in $H_1$ all terms of degree $2$ in the actions $p$ and $K$
in the angles $q$, and in $H_2$ all terms of degree $3$ in the actions
$p$ and $2K$ in the angles $q$. The Birkhoff normalization produces a
Hamiltonian of the form
$$
H = \langle\omega,p\rangle + Z_1(p) + \ldots + Z_r(p) + \Fscr_{r+1}(p,q),
$$
where $\Fscr_{r+1}$ denotes the term of degree $r+2$ in the actions
$p$ and $(r+1)K$ in the angles, i.e., the first term of the remainder.
With a suitable choice of $K$, this considerably reduces the number of
coefficients in the expansions thus enabling us to perform the
calculation on a workstation.  We emphasize however that the algorithm
is a general one so that in principle it can be applied to the full
Hamiltonian or, better to a Hamiltonian obtained by removing all
coefficients which are very small and will likely not produce big
coefficients (due to the action of small denominators) during the
calculation of the Birkhoff's normal form.  The rest of the
calculation closely follows the discussion in section~\secref{3}, so
we come to illustrating the results.  We performed the calculation
with two different values of $K$, as given in the following table.
$$
\vcenter{\openup1\jot\halign{
\hfil$\displaystyle{#}$
&\qquad\hfil$\displaystyle{#}$
&\qquad$\displaystyle{#}\hfil$
\cr
K  &  r   &   {\rm \#\ of\ coefficients}
 \cr
4  &  5   &    2\,494\,000         
\cr
6  &  4   &    3\,380\,000         
\cr
}}
$$

This shows in particular the dramatic increase of the number of
coefficient in the remainder $\Fscr_{r+1}$ (third column), which
imposes strong constraints on the choice of the normalization order
$r$.

A quite natural objection could be raised here.  Since the most
celebrated resonance of the SJS system (i.e., the mean motion
resonance $5:2$) has trigonometrical degree~$7\,$, it seems that some
of the main resonant terms are neglected because of our choice of
$K\,$.  This is actually not the case, due to a technical element that
we have omitted in the previous section in order to make the
discussion simpler.  Our sequence of transformations includes a
unimodular linear transformation on the angles, and so also on the
frequencies.  The action on the frequencies changes the resonance
$5:2$ into a $3:1$ one, which is of order $4$.  Thus, setting $K\ge 4$
as we did throughout all our calculations is enough in order to
include the main resonant terms.  The interested reader will find a
detailed discussion of this point in sect.~3
of~\dbiref{Locatelli-2007}.

\figure{std_omo}{\vbox{\openup1\jot\halign{
\hfil$\displaystyle{#}$\hfil
\cr
 \rlap{\kern 2.5pt\raise 5pt\hbox{\bf(a)}}\ignorespaces
  \psfig{file=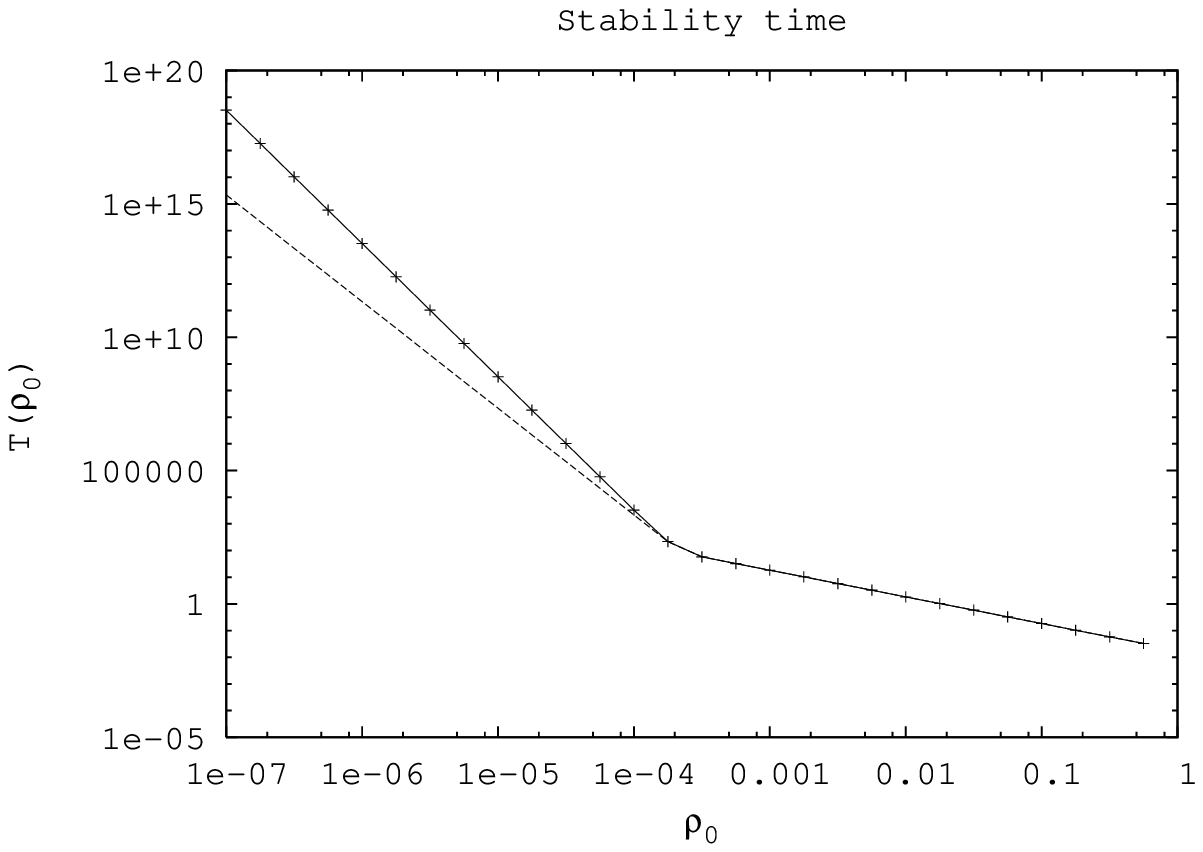,width=13 truecm}\quad
\cr
  \rlap{\kern 2.5pt\raise 5pt\hbox{\bf (b)}}\ignorespaces
    \psfig{file=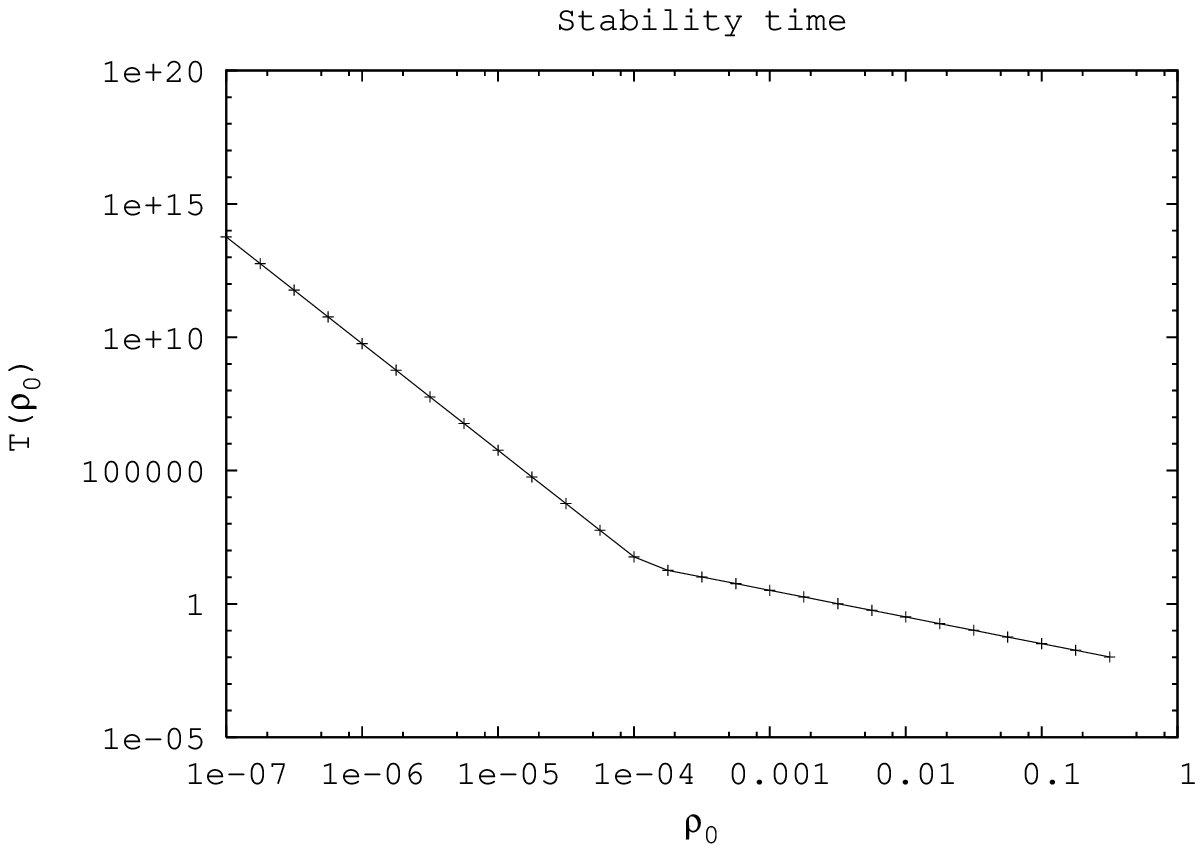,width=13 truecm}
\cr
}} }{The estimated stability time. (a)~results for $K=4$ and Birkhoff
normalization order $5$ (crosses) and $4$ (dashed curve).  (b)~results
for $K=6$ and Birkhoff normalization order $4$.}

Let us now come to the results.  In panel~(a) of fig.~\figref{std_omo}
we report the results for $K=4$.  The crosses give the estimated
stability time for the Birkhoff's normal form at order~5; the dashed
line gives the estimated time when the Birkhoff's normal form is
truncated at order 4, thus showing how relevant is the improvement
when a single normalization order is added.

We can now come back to the estimate about the escape time
$T=T(\rho_0)\,$. Looking at panel~(a) of fig.~\figref{std_omo}, one
can remark that we have an estimated stability time of
$10^{10}$~years, that is approximately equal to the age of the
universe, for a neighbourhood of initial conditions of radius
$10^{-5}$ in actions.  

It may be noted that the stability curves exhibit a sharp change of
slope around $\rho_0\sim 10^{-4}$. This is because the optimal
normalization order increases when the radius is decreased.  Actually,
further changes of slope should be expected for smaller values of
$\rho_0$, but due to computational limits such changes can not appear
in our figure, because the optimal order exceeds the actual order of
our calculation.  Thus, our estimate of the stability time should be
considered as a very pessimistic lower bound.

Moreover, the behaviour of the plots in fig.~\figref{std_omo}, clearly
shows that the estimate of the escape time can be substantially
improved if smaller values of the radius $\rho_0$ can be considered.
Recalling that our estimate of the size of the neighbourhood in action
variables is calculated from the discrepancy among different sets of
JPL's DE data, we may affirm that our neighbourhood roughly covers
such a width, which is tabulated in Standish's paper quoted above.

If we try a better approximation of the Hamiltonian, setting $K=6$,
then we are forced to stop the Birkhoff's normalization at order 4,
thus making the results definitely worse.  The data for the estimated
stability time are plotted in panel~(b) of fig.~\figref{std_omo}.  One
sees that the estimate becomes comparable with the age of the universe
only in a neighbourhood of initial conditions
slightly larger than $10^{-6}$.  However, if we compare the curve in
panel~(b) with the dashed curve in panel~(a) we see that we shall
likely get substantially better results if we could compute the normal
form at order 5.

Thus, our rough approximation gives results which apply to a set of
initial data for the SJS system which is of the same order of
magnitude as the uncertainty in JPL's data.  We also emphasize that
our evaluation of $\rho_0$ is based on observational data which are
presently older than $25$~years; we expect that this is quite
pessimistic with respect to the features of more recent JPL's DE.

\section{6}{Conclusions}
We have developed an effective method to compute the Kolmogorov's
normal form for the problem of three bodies, and have successfully
applied it to the SJS problem, using the data of our Solar System.
Next, we have shown that a calculation of the stability in
Nekhoroshev's sense of orbits with initial point in the neighbourhood
of the torus is possible, at least if one accepts to make some strong
truncation in the expansion of the Hamiltonian.  A rather strong
truncation allows us to get an estimated stability time comparable
with the age of the universe in a neighbourhood of the torus that will
likely contain the actual initial data of the SJS system.

The natural question is whether such results will remain valid if we
add more and more terms in the Hamiltonian, thus making our
approximation better.  Answering such a question is presently beyond
our limits, but in our opinion deserves to be investigated.  Some
improvement may be obtained by using more computer power, e.g., by
performing our calculation on a cluster of computers.  However,
substantial improvements require also a refinement of our analytical
techniques in order to be able to evaluate the error induced by our
truncations, thus allowing us to introduce better computation schemes
and to evaluate the reliability of our approximations.

\references

\bye